\documentclass[12pt]{article}
\usepackage{epsfig,latexsym,makeidx}
\usepackage{amsmath,amsfonts}
\usepackage{subfigure}
\usepackage{amssymb, comment,wrapfig}
\usepackage{epstopdf, caption}
\usepackage{pstricks, nicefrac}
\usepackage{chemarr, color}
\usepackage[top=1in, bottom=1in, left=1in, right=1in]{geometry}

\newcommand{\beq}{\begin{equation}}
\newcommand{\eeq}{\end{equation}}
\newcommand{\ben}{\begin{eqnarray}}
\newcommand{\een}{\end{eqnarray}}
\newcommand{\bea}{\begin{array}}
\newcommand{\eea}{\end{array}}
\newcommand{\bef}{\begin{figure}}
\newcommand{\eef}{\end{figure}}

\date{ }

\begin{document}

\title{\large \bf Chapter 9 TISSUE ENGINEERING} 
\maketitle

%\vskip 100pt

\setcounter{section}{+9}
\subsection{Introduction}
Tissue Engineering (TE) is an interdisciplinary field dealing with the principles of engineering and life sciences toward the development of biological substitutes that restore, maintain, or improve tissue function or a whole organ [52]. Currently, TE is emerging as an invaluable field of study and is one of the most promising fields for the next century.  The ultimate goal is to develop powerful new therapies for biological substitutes Ð that will successfully restore the structural and functional disorders of human and animal system.

Available standard treatments for organ or tissue loss include transplantations of the same organ or organs from healthy donors, surgically reconstructed with synthetic implants in most of the cases, and the use of mechanical devices such as kidney dialysers, 
\begin{wrapfigure}{r}{0.4\textwidth}
  \centering
  \vskip -20pt
  \includegraphics[width=0.4\textwidth]{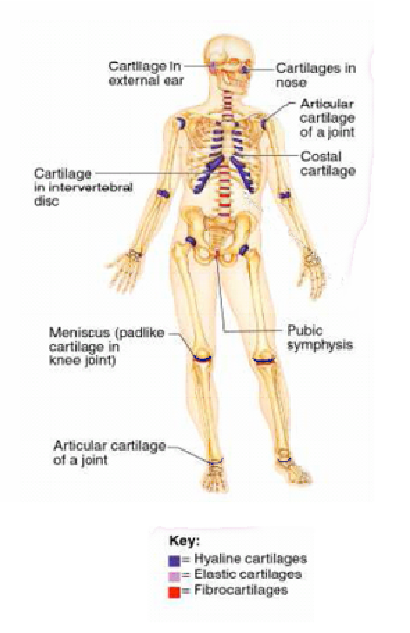}
\caption*{Fig 9.1: Possible locations of different cartilages in human body ({\it Source: [1]}).}\label{fig:Fig1}
\end{wrapfigure}
ventricular assist device (VAD) etc [14]. Of course, these treatments can save lives and decrease morbidity. However, each has its significant drawbacks and most are associated with substantially decreased quality of patientÕs life.  Although these procedures (treatments) temporarily save countless lives, unfortunately these techniques are not perfect.

Tissue engineering procedures offer a new hope in modern medical science. By using patientÕs own cells to produce bioengineered tissues and organs that function like their natural counterparts, TE may offer a better solution for the repair and reconstruction of damaged tissue and organs. At present, the only successfully engineered tissue that has been made available to people is skin. Major success in this direction came in 1984 when Robert Langer, a chemical engineer at MIT and Joseph Vacanti, a medical surgeon at Harvard, first time jointly came up with an idea of duplication of bodyÕs production tissues by developing  a biodegradable scaffold on which skin cells could be grown using fibroblasts. After rigorous and extensive experiments with different techniques they discovered that collagen and fibroblasts can arrange themselves in a compact dense form under specific physical conditions that is exactly similar to the inner layer of the skin.  Seeding keratinocytes on this layer leads to the formation of upper layer of the skin. This is how the artificial skin was developed by TE methods. The important point of this artificial skin is that it has no immunogenic cells, therefore question of rejection does not arise when implemented on patientÕs body. Artificial skin are now made available commercially in large quantities. The next major step towards the development of artificial organ using tissue engineering approach is the cartilage. It is believed that the cartilage has relatively simple composition and structure.

%%%%%%%%%%%%%%%%%%%%%%%%%%%%%%%%%%%%%%%%%%%%%%%%%%
\setcounter{section}{+9}
\subsection{Cartilage Tissue Engineering} \label{sec:intro}

\subsubsection{Introduction}

\begin{wrapfigure}{r}{0.45\textwidth}
  \centering
  \includegraphics[width=0.45\textwidth]{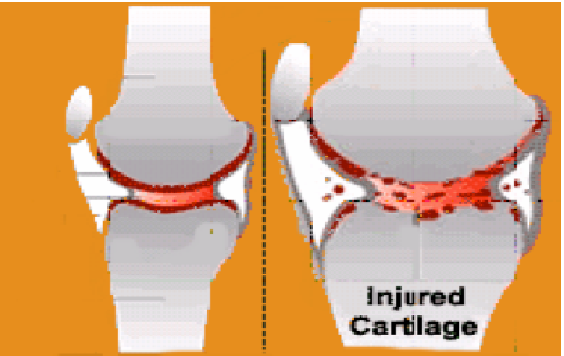}
 \caption*{Fig. 9.2: A damaged knee ({\it Source: [66]}).}\label{fig:Fig2}
\end{wrapfigure}
Articular cartilage is a living tissue but of very limited capacity to repair itself. In fact, any damage to the articulating surface (due to injury, disease, and/or genetic disorders) tends to spread allowing the bones to rub directly against each other and resulting in increasing joint pain and loss of joint movement. It can also lead to osteoarthritis (a slow degeneration of articular cartilage) (see Fig. 9.2). It is the most common form of arthritis in Australia, about 10\% of the total population [61]. In United States alone, more than one million people annually suffer from osteoarthritis, a degenerative joint disease caused by damage to articular cartilage [34]. Damage to the articulating surface can occur due to compression of heavy load or when angular and/or shear forces are applied on it. With the prevalence of cartilage-related disease (e.g., Osteoarthritis), injuries and cost of treatment on the rise, the Osteoarthritis Research Society International (OARSI) urges new treatment options [34]. An overview of a healthy and damaged knee is given in Fig. 9.2. 

The ability of cartilage to withstand enormous compressive loads (many times the body weight) without being crushed, is attributable to the multi-phasic nature of the tissue. Mature articular cartilage contains approximately 5\% of its volume as cells (also known as the growth cells or chondrocytes), and 95\% as extracellular matrix (ECM). From an engineering standpoint, ECM is a porous, viscoelastic material consisting of three principal phases (a phase represents all of the chemical compositions with similar physical properties): 1) the solid phase constitutes about 30\% of the total weight, and contains collagen (60\%), proteoglycans or PG (25\%) and wide range of matrix proteins (15\%), 2) the fluid phase (also known as the synovial fluid), which constitutes on average about 70\% of the total weight, for normal tissues, and 3) an ion phase, consisting of many ionic species of dissolved electrolytes (e.g., Na$^+$, Ca$^{++}$, Cl$^-$). These constitute less than 1\% of the net weight. The biofluid mechanics of cartilage growth and degradation relies on this multi-phasic nature of the tissue.

\begin{wrapfigure}{r}{0.45\textwidth}
\vskip -10pt
  \centering
  \includegraphics[width=0.45\textwidth]{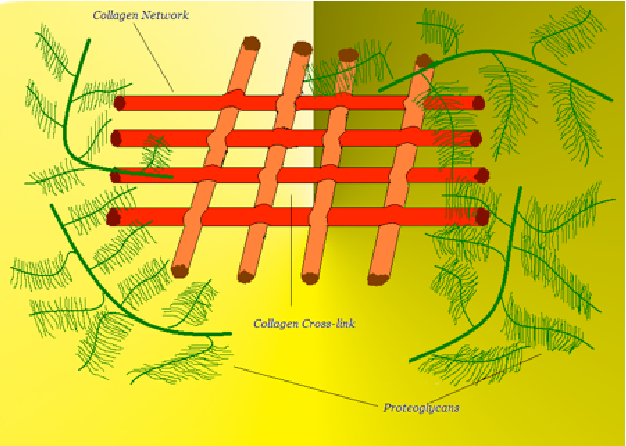}
 \caption*{Fig. 9.3: Distribution of proteoglycan \& collagen in articular cartilage ({\it Source: [79]}).}\label{fig:Fig3}
\end{wrapfigure}

%%%%%%%%%%%%%%%%%%%%%%%%%%%%%%%%%%%%%%%%%%%%%%%%%%
\subsubsection{Collagen} \label{subsec:collagen}

The collagen fibrils are densely packed polymer strands with a high resistance against fluid flow. They are the dominant structural components of the solid matrix and helps in retaining the shape of the cartilage when compressed. Collagens are basically made off three polypeptide chains arranged as triple helix long its length. There are two different types of collagens, known as fibrillar and non-fibrillar collagens. Depending upon structural and functional characteristics collagen has been classified as 12 different categories. Hyaline cartilage (articular cartilage) mainly consists of type-II and type-XI fibrillar collagens where the percentage of type-II collagen is much higher.

The architecture of articular cartilageÕs connective tissues is very complicated. Collagen fibrils in
the superficial-most region (known as the tangential zone) of the cartilage layer are densely packed and oriented parallel to the articular surface. This collagenous membrane has a relatively low PG content and a lower permeability to fluid flow, which is important in providing for a barrier of high resistance against fluid flow when cartilage is compressed. In the middle or transitional zone of the cartilage, the collagen fibers are larger and are either randomly or radially orientated. In the deepest zones, i.e., the zone nearest to the calcified cartilage and sub-chondral bone interface the collagen fibers are larger and form bundles that are oriented perpendicular to the calcified/bony interface. The way how the collagen and proteoglycan/glycosaminoglycans (GAG) molecules are arranged inside the articular cartilage, is shown in Fig. 9.3.

%%%%%%%%%%%%%%%%%%%%%%%%%%%%%%%%%%%%%%%%%%%%%%%%%%
\subsubsection{Proteoglycan} \label{subsec:PG}

\begin{wrapfigure}{r}{0.4\textwidth}
%\vskip -20pt
  \centering
  \includegraphics[width=0.4\textwidth]{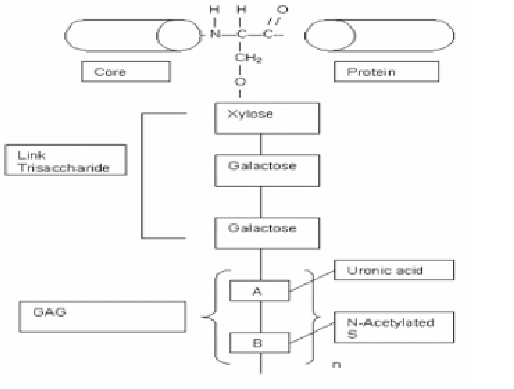}
 \caption*{Fig 9.4: A GAG molecule attached to the core protein in proteoglycan}\label{fig:Fig4}
\end{wrapfigure}
PG are macromolecules consisting of a protein core to which are attached short chains of negatively charged glycosaminoglycans (GAGs). The primary GAGs associated with PGs in cartilage are chondroitin 4-sulfate, chondroitin 6-sulfate and keratan sulfate (Fig. 9.4). It is available in all connective tissues but its structure varies with the type of tissue. Hyaline cartilage contains mainly aggregating proteoglycans.

Each PG-associated negative charge on the polymer chain (known as the fixed charge, FC), requires a mobile counter-ion (e.g., Na$^+$) dissolved within the interstitial fluid in the cartilage gel to maintain its electro-neutrality. This effect gives rise to an imbalance of mobile ions across the gel interface. The excess of mobile ions colligatively yields a swelling pressure, known as the osmotic pressure, while the swelling pressure associated with FCs is known as the Donnan pressure. Changes in this internal swelling pressure, arising from altered ion concentrations of the external bath, or changes in the fixed charges result in changes in tissue dimensions and hydration. However, this swelling pressure is balanced by tensile forces generated in the collagen network. This effect is due to the presence of covalent cross-links within the polymer matrix. An increase in cartilage tissue hydration, governed by the density and the nature of FCs on the PGs as well as the density of the mobile counter ions in the interstitial fluid are the earliest signs of articular cartilage degeneration during osteoarthritis.

%%%%%%%%%%%%%%%%%%%%%%%%%%%%%%%%%%%%%%%%%%%%%%%%%%
\subsubsection{Chondrocyte} \label{subsec:Chon}

\begin{wrapfigure}{r}{0.4\textwidth}
\vskip -43pt
  \centering
  \includegraphics[width=0.4\textwidth]{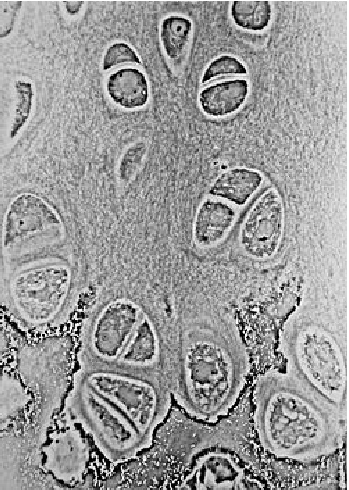}
 \caption*{Fig. 9.5: Light micrograph of hyaline cartilage showing its chondrocytes and organelles, lacunae and matrix ({\it Source: Wikipedia}).}\label{fig:Chon}
\end{wrapfigure}
Chondrocytes are mature cells found in cartilage. They make up the cellular matrix of cartilage, performing a number of functions within the tissue, including production and maintenance of the cartilaginous matrix and facilitating the exchange of fluids through the gelatinous layers. Fully mature chondrocytes tend to be round, and they may cluster together in small groups within the network of the cartilage (Fig. 9.5).

The progenitors of these cells arise in the bone marrow, in a form of stem cell. Stem cells are capable of differentiating into several different types of cell, depending on the need. When they differentiate into cartilage cells, they start out as chondroblasts, actively producing secretions of chondrin, the primary substance in cartilage, to build and repair the tissue. Once a chondroblast becomes totally surrounded, it is a mature chondrocyte. These cells can be found in small gaps within the cartilage known as lacunae.

Chondrocytes are not capable of cell division. They can produce secretions to support and repair the cartilage matrix, and as discussed above, the facilitate the exchange of materials between the cartilage and the surrounding material. Depending on what type of cartilage a cell is in, it may have a slightly different composition; elastic, hyaline, and fibrocartilage are all unique, designed to meet various needs within the body.

%%%%%%%%%%%%%%%%%%%%%%%%%%%%%%%%%%%%%%%%%%%%%%%%%%
\subsection{Mathematical model of the ECM} \label{sec:model}

Mathematical models are essential for understanding systems (like cartilage) with many interacting components (e.g. the collagen network, proteoglycans, interstitial fluid and ionic solutions present in cartilage), since the complexity of the interactions taking place overwhelms the ability of the human intuition to make accurate predictions of their behavior. Comparisons of the models' quantitative predictions with experimental results allows us either to confirm that the biological hypotheses regarding these interactions (which are encoded in the model) are correct or, equally usefully, identify that some of them are in error. Once thus validated, the model will ultimately help design treatments related to cartilage loss and osteoarthritis by predicting this outcome of proposed interventions, and focusing efforts on those which appear most promising.

In this chapter, an account of the equilibrium for electro-chemical properties of poly(ethylene gylcol) gel (PEG) containing negatively charged chondrotin sulfate (ChS) strands, immersed in water with several different ions: H$^+$, Na$^+$, Cl$^-$ is demonstrated. This material closely mimics the electro-chemical properties of articular cartilage. We view this mixture as a multi-component material, including solvent particles, polymer and particles of several ionic species. The polymer is assumed to be made up of two types of monomers the uncharged units (PEG segment of the polymer) and those that are the charged ones (ChS segment of the polymer). Only the charged units participate in the binding reactions, carry a double negative charge and are denoted as M$^{2-}$. The positively charged ions in the solvent are Hydrogen (H$^+$) and Sodium (Na$^+$). The negatively charged ions could include Hydronium (OH$^-$) and Chloride (Cl$^-$). Because the negatively charged ions are assumed to be not involved in any binding reactions with the gel, acting only as counterions to positive charges, we identify these ions by the name Chloride. The binding reactions of the positively charged ions with the monomers are: \vskip -10pt
 \ben
&&(a) \hspace{0.25cm} {\rm M^{2-}}+{\rm H^+} \xrightleftharpoons[\text{k$_{-h}$}]{\text{k$_h$}} {\rm MH^-} , ~~(b) \hspace{0.25cm} {\rm M^{2-}}+{\rm Na^+} \xrightleftharpoons[\text{k$_{-n}$}]{\text{k$_n$}} {\rm MNa^-}, ~~(c) \hspace{0.25cm} {\rm MH^{-}}+{\rm H^+} \xrightleftharpoons[\text{k$_{-h2}$}]{\text{k$_{h2}$}} {\rm MH_2}, \nonumber \\
&&(d) \hspace{0.25cm} {\rm MNa^{-}}+{\rm Na^+} \xrightleftharpoons[\text{k$_{-n_2}$}]{\text{k$_{n_2}$}} {\rm MNa_2}, ~~(e) \hspace{0.25cm} {\rm MH^-} +{\rm Na^+} \xrightleftharpoons[\text{k$_{-hn}$}]{\text{k$_{hn}$}} {\rm MHNa}. \label{eqn:chem}
\een
We assume that all the binding sites/charge sites are identical and the binding affinities for the different ions are different. The species M$^{2-}$, MH$^-$, MNa$^-$, MH$_2$, MNa$_2$ and MHNa are different monomer species of the charged type, all of which move with the same polymer velocity.  The ion species are freely diffusible, but because they are ions, their movement is restricted by the requirement to maintain electroneutrality.

Suppose we have some volume $V$ of a mixture comprised of $k$ types of particles each with particle density (number of particles per unit volume) $n_j$, and particle volumes $\nu_j$, $j = 1,\cdots, k$. From now on we will denote the quantities with subscript ``1'' related to PEG monomer species, subscript ``2'' related to ChS monomer species and subscript ``3'' related to solvent molecule.  The subscript (p,s) denotes polymer and solvent phase, respectively. The volume fraction for each of the polymer species ($\theta_1, \theta_2$) and the solvent ($\theta_3 = \theta_s$) are $\theta_i=\nu_i n_i$ ($i = 1, 2, 3$) respectively and $\theta_p=\theta_1+\theta_2$. Suppose the ChS monomers constitute a fraction `$\alpha$' of the total number of monomers, $n_p$,  (i.e., $n_2 = \alpha n_p$, $n_1 = (1-\alpha) n_p$). Assuming that the ionic species do not contribute significantly to the volume (i.e., we take $\nu_j=0$, $j=4,\cdots, k$), conservation of total volume implies $\theta_1+\theta_2+\theta_3 = 1$. Further, in subsequent calculations we make an assumption that the particle density of the ions is insignificantly small compared to the particle density of polymer and solvent, i.e. $\sum_{j\ge4} n_j \ll n_s+n_p$. Hence, $\phi_s = \frac{n_s}{n_p+n_s}$, $\phi_p = \frac{n_p}{n_p+n_s}$, $\phi_j = \frac{n_j}{\sum_{i\ne 1,2}n_i}$ for $j\ge 4$ (assuming that the ions are dissolved in the solvent), are the polymer, solvent, and ion species per total solvent particle fractions, respectively.

%%%%%%%%%%%%%%%%%%%%%%%%%%%%%%%%%%%%
\subsubsection{Chemical potential} \label{CP}

This section outlines the calculation of the free energy and the chemical potential for this multi-species material. As above, we suppose there are $k$ different kinds of particles, with particle volumes $\nu_i$ and particle numbers $n_i, i = 1, \dots, k$. To calculate the chemical potentials, $\mu_j$, we use the Gibb's free energy
\beq 
G = -k_BTS + U + PV,\label{eq:gibbs_fe} 
\eeq 
where $U$ is internal energy, $S$ is entropy, $T$ is temperature, $k_B$ is Boltzmann's constant, $P$ is pressure, and $V=\sum_j\nu_j n_j$ is the total volume of the system. However, we assume that the  volume occupied by species $j=4, \cdots, k$ (which are the ions dissolved in the solvent) is small compared to that of the monomer plus solvent volumes $\nu_p n_p+\nu_sn_s$  and so  take $V = \nu_p n_p+\nu_s n_s$. The chemical potential are given by
\beq
\mu_j = \frac{\partial G}{\partial n_j} = -k_BT\frac{\partial S}{\partial n_j} + \frac{\partial U}{\partial n_j} + \nu_j P = \mu^S_j + \mu^I_j + \nu_j P, \label{eq:chem_pot}
\eeq
where $\mu^S_j$, $\mu^I_j$ are the contribution due to entropy and internal energy respectively. In the next two sections these two contributions to the chemical potential are outlined.

%%%%%%%%%%%%%%%%%%%%%%%%%%%%%%%%%%%%
\subsubsection{Entropic Contributions to Chemical Potentials ($\mu^S_j$)} \label{CPEntropy}

The entropy of the system is defined as 
\beq 
S = \sum n_i\omega_i,
\eeq
where $\omega_i$ is the entropy per particle for the $i^{th}$ particle. Using standard counting arguments given by Flory [23], for single-molecule species,
\beq \omega_j = -\ln(\phi_j) \qquad j \ge 3 
\eeq
The PEG and ChS chains exhibit permanent cross-link bonds (i.e., covalent bonds), and we calculate the per-particle entropy of these two species using the Doi's rubber elasticity theory [20]
\beq
\omega_i = -\frac{3k_i}{2 N_i} \Big[ (\phi_i)^{-2/3} - 1 \Big], \qquad i=1,2, \label{eq:rubberElas}
\eeq
where $i=1, 2$ corresponds to PEG and ChS polymer species respectively. The corresponding particle fractions are $\phi_1 = (1-\alpha) \phi_p$ and $\phi_2 = \alpha \phi_p$. $k_i$ and $N_i$ are the fractions representing the number of cross-linked monomers per total number of monomers in one chain, and the number of monomers per chain, respectively. Finally, the expressions for the entropic chemical potentials are given in the article by Sircar {\it et al.} [82] and are not included here for brevity.

%%%%%%%%%%%%%%%%%%%%%%%%%%%%%%%%
\subsubsection{Internal Energy Contribution to Chemical Potentials ($\mu^I_j$)} \label{CPIntEnergy}

The internal energy consists of two contributions, long range electrostatic interactions and short range (nearest neighbor) interactions. The long range electrostatic interactions have energy
\beq
U_e = \sum_j z_jn_j\Phi_e,
\eeq
where $z_i$ is the charge on the $i^{th}$ ionic species ($z_1 = z_2 \equiv z_p$ is the average charge per monomer), and $\Phi_e$ is the electric potential.

To calculate the short range interaction energy for the polymer and solvent, we assume that each of the $n_T$ (= $n_p + n_s$) particles have $z$ neighboring interaction sites (called the coordination number). Of the total of $n_p$ monomers, $k_1 n_1$ and $k_2 n_2$ of them are the cross-linked PEG and ChS monomers, respectively. Since the cross-linked particles are pair-wise connected, we treat each of these pair as a single species.

The different species with their pairwise interactions (i.e., the cross-linked species, x, the uncross-linked species, u, and the solvent particles, s) are shown in Fig. 9.6. Following assumptions are made while calculating the per-particle interaction energy, $U$: (1) all cross-links are assumed identical, (2) the cross-links are covalent permanent bonds, (3) the polymers have long chains and hence the end-effects are neglected, and (4) different polymer species have the same interaction energies if they are of the same type (e.g., uncross-linked PEG and uncross-linked ChS monomers have the same interaction energy). The cross-linked particle pair and the uncross-linked monomer have 2$z$-6 and $z$-2 free interaction sites based on their position in the middle of the polymer chain, respectively (Fig. 9.6(a) and Fig. 9.6(b)). The solvent particles have $z$ free interaction sites (Fig. 9.6(c)). 
\begin{figure}[htbp]
\centering
\includegraphics[scale=0.18]{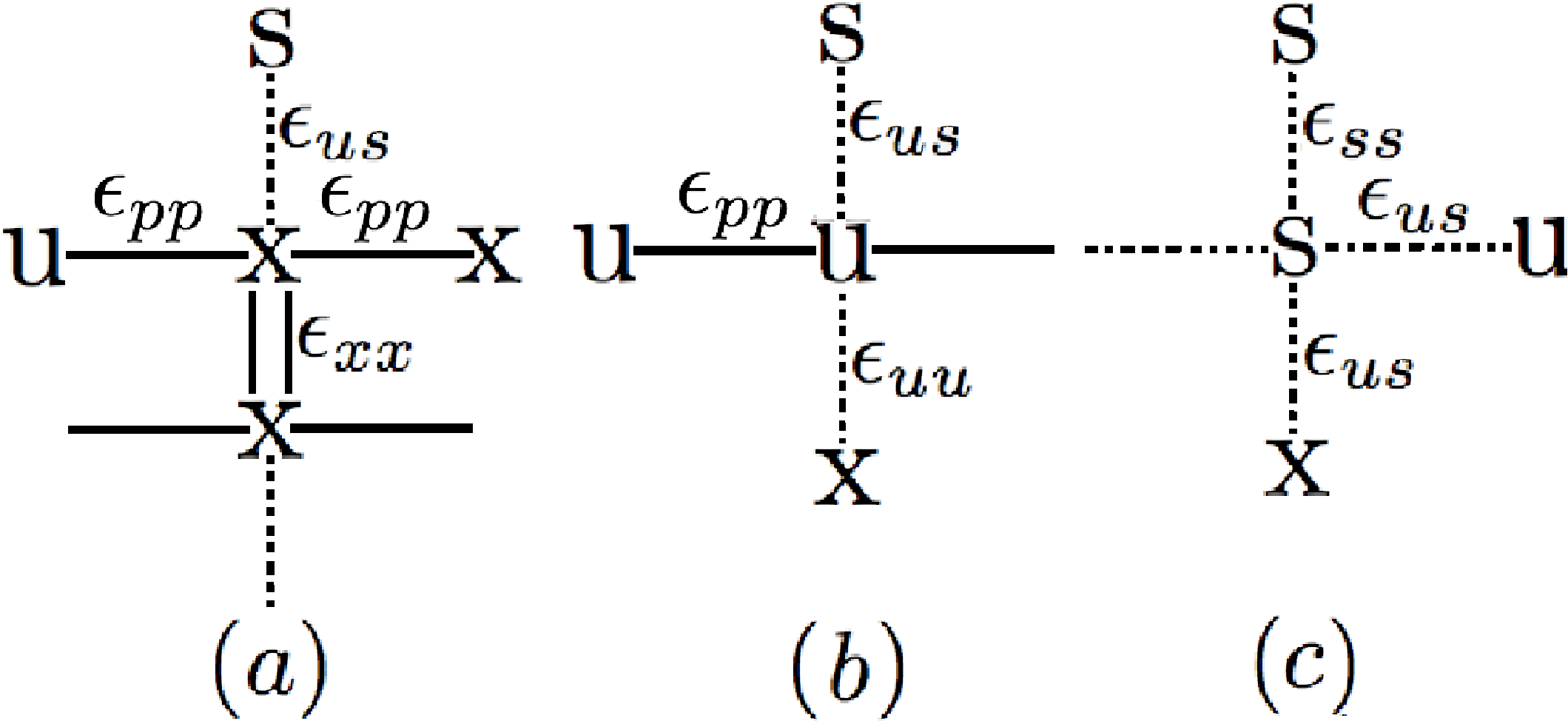}
\caption*{Fig. 9.6: Pairwise interactions of the different species and their associated interaction energies for: (a) cross-linked monomers, x,  (b) uncross-linked monomers, u, and (c) solvent, s. Double solid lines denote cross-linking, single solid lines denote interactions within a polymer chain while the other types of interactions are denoted by dashed lines. The nearest neighbor interactions are assumed identical for both the PEG and ChS monomer species.} \label{fig:Fint}
\end{figure}

The interaction energies for crosslinked, uncrosslinked polymer particles, and solvent particles are calculated using standard mean-field arguments. For example, the interaction energies for cross-linked particles, $F_x$ (Fig. 9.6a), are calculated by noting that there are either $\frac{n_1}{2}$ pairs of only PEG monomers, $\frac{n_2}{2}$ pairs of only ChS monomers or $\frac{n_1+n_2}{2}$ pairs of both PEG and ChS monomers, respectively. The corresponding probabilities of forming cross-links are $k^2_1, k^2_2$ and $k_1k_2$. For each of these pairs, two neighboring sites are occupied by the covalent cross-link, four neighboring interactions are the monomer-monomer interactions within the polymer strands, and the remaining $z-6$ interactions are distributed to monomer-monomer, with probability $\frac{n_p}{n_T}$ and  monomer-solvent, with probability $\frac{n_s}{n_T}$. Hence,
\beq
\frac{F_x}{k_B T_0} = \left( \frac{n_1}{2} k^2_1 + \frac{n_2}{2} k^2_2 + \frac{n_1 + n_2}{2} k_1 k_2 \right) \left(2\epsilon_{xx}+(2z-6)E_{us}+ 4\epsilon_{pp}\right), \label{eq:Fx}
\eeq
where $E_{us} =  \epsilon_{uu} \frac{n_p}{n_T} + \epsilon_{us}\frac{n_s}{n_T} $. Similar arguments are made to calculate the interaction energies for uncrosslinked particles ($F_u$) and solvent particles ($F_s$). The total per particle interaction energy is
\begin{align}
U &= \frac{1}{2(n_m+n_s)}(F_x+F_u+F_s)\nonumber
 \\
 &= k_B T_0 \left[\chi \phi_p\phi_s + \mu_0^s \phi_s+\mu_0^p\phi_p + \frac{z}{2}\epsilon_{us} \right],\label{eq:29}
\end{align}
where the coefficients, $\chi, \mu_0^p, \mu_0^s$, are referred as the Flory interaction parameter and the chemical potentials of pure polymer and solvent species respectively [23]. These are functions of the particle fraction of the ChS strands ($\alpha$), the interaction energy associated with the uncross-linked monomer-monomer interaction ($\epsilon_1$), solvent-solvent interaction ($\epsilon_2$), covalent cross-links interaction ($\epsilon_3$) and monomer-monomer bond within a polymer strand ($\epsilon_4$). The factor $\frac{1}{2}$ in the per-particle interaction energy in Eqn.  (\ref{eq:29}) is to correct for double counting. The corresponding contributions to chemical potentials ($\mu^I_j$) are calculated via Eqn. (\ref{eq:chem_pot}).

%%%%%%%%%%%%%%%%%%%%%%%%%%%%%%%%%%%%
\subsubsection{Interface conditions} \label{IC}

Under no-loading conditions, there is a free moving-edge to the gel on one side of which (inside the gel) $\theta_p = \theta_p^-$, $\theta_s = \theta_s^-$, and on the other side of which (outside the gel)  $\theta_p^+ = 0$, $\theta_s^+ = 1$,  there are  interface conditions, 
\beq
\frac{1}{\nu_p} \mu_p^-  {\bf n}=0, \label{eq:edge_Conditions_3}
\eeq
for the polymer, and
 \beq
\frac{1}{\nu_s}(\mu_s^+ - \mu_s^-) {\bf n}=0, \label{eq:edge_Conditions_4}
 \eeq
for the solvent. {\bf n} is the normal to the free surface. The interface conditions are derived using the standard variational arguments to minimize the rate of work, which constitutes the viscous rate of energy dissipated within the polymer and the solvent, energy dissipation rate due to the drag between solvent and polymer and between solvent and ion species particles as well as the rate of work required against the chemical potential, $\mu_j$. These interface conditions contain the contribution to the swelling pressure via the entropy, the Donnan potential and the net osmolarity (or the swelling pressure coming from the difference between the concentrations of ions dissolved in the gel and those dissolved in the bath, outside the gel). The derivation details can be seen in [46, 47, 82].

 %%%%%%%%%%%%%%%%%%%%%%%%%%%%%%%%
\subsubsection{Ionized-species chemistry} \label{Chem}

We use the law of mass action to describe the chemical reaction. Under the assumption of fast chemistry and diffusion, the forward and the back reaction rates in Eqn. (\ref{eqn:chem}) are balanced. Those relations provide us with quasi-equilibrium concentration of the monomer species (e.g., $m=[\text{M}^{2-}], y=[\text{MH}^-], v=[\text{MNa}^-], w=[\text{MH}_2], x=[\text{MNa}_2]$ and $q=[\text{MHNa}]$). Similarly, the quasi-equilibrium concentration of the ion species ($n = [{\rm Na^+}]$, $h = [{\rm H^+}]$ and $c_l = [{\rm Cl^-}]$) reduces into
\beq C = C_b e^{-z_C \Psi_e}, \label{eq:ion_bal3c} 
\eeq
where $C = h, n, c_l$, $z_n= z_h=1$ and $z_{c_l} = -1$ and the subscript `b' denotes the corresponding bath concentrations (or concentrations outside the gel). The electrostatic potential, $\Psi_e$, is determined by the electro-neutrality constraint inside the gel, namely,
 \beq  (n + h-c_l)\theta_s + z_p m_T=0, \label{eq:el_bal} \eeq
where $z_p$ is the average charge per monomer. This charge depends on the residual charge of the unbound and the bound ChS monomers, $z_p$, given by
\beq z_p m_T = -(2 m + y + v). \label{eq:zp}\eeq
From Eqns (\ref{eq:el_bal}, \ref{eq:zp}), we can calculate the Donnan potential, $z_p \Psi_e$.
This potential is zero outside the gel (since there is no polymer in that region). 

%%%%%%%%%%%%%%%%%%%%%%%%%%%%%%%%%%%%%%%%%%%%%%%%%%
\subsection{Experimental methods} \label{sec:exp}

{\em In vitro} experiments provide the most intuitive mechanism for cartilage swelling-deswelling kinetics. Recent experiments have confirmed that changes in external ion concentration and pH affect not only the mechanical properties of cartilage, but also the biosynthetic capabilities of chondrocytes. The experimental outcomes of Beekman {\it et al.} [7], Freed {\it et al.} [25], Kisiday {\it et al.} [49], Regan {\it et al.} [73], Stading and Langer [84], Vunjak-Novakivic {\it et al.} [96] and Wilson {\it et al.} [99], all share a common feature: collagen accumulation in engineered cartilage tend toward a constant value (steady state) with respect to time. A similar type of result was obtained by Buschmann [16] for GAG molecule accumulation in engineered cartilage constructs.

\begin{wrapfigure}{r}{0.4\textwidth}
\vskip -15pt
  \centering
  \includegraphics[width=0.4\textwidth]{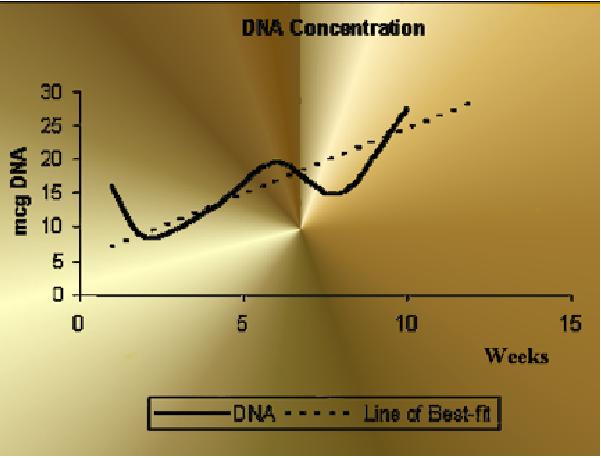}
 \caption*{Fig 9.7: DNA concentration {\it Source: [99]}}\label{fig:Fig5}
\end{wrapfigure}
According to the results on DNA content of engineered cartilage constructs obtained by Wilson {\it et al.} [99] (Fig. 9.7), there is an increasing trend of cell population growth that means the cell population has no control (i.e., it keeps on increasing unboundedly), but this is not the case in nature. Hence, overall experimental observations suggests that ECM deposition in cell-polymer construct of engineered cartilage heading towards a steady state but there is no concrete evidence that chondrocytesÕ population does the same.

Our experimental validation of the model, described previously, is performed by running a subset of {\em in vitro} experiments in tandem with the proposed computational experiments described below. We construct ionic gels by co-polymerizing methacrylated chondroitin sulfate and poly(ethylene glycol) dimethacrylate in deionized water (distilled water). PEG was chosen since it can be used as a neutral, cross-linked, 3-D scaffold, a system that promotes the deposition of proteoglycans and collagen molecules and emulates the mechanical strength, load bearing capabilities and resilience of cartilage tissue. The ChS component serves two purposes: to mimic the biochemical environment of cartilage (since it is the main component of proteoglycans) and to introduce fixed negative charges into the network. These polymer gels are then allowed to reach equilibrium in deionized water (e.g., 24 hours) and then placed into appropriate solution (e.g., with variable pH, salt concentration, etc.) to reach a new equilibrium.

 %Based on our preliminary results (Fig. \ref{fig:Fig2}), we expect excellent correlation between experimental and numerical results. 

To quantify the local configuration of the polymer gels, the counter-ion concentrations and the osmolarity inside these gels, the following in-vitro experiments were performed and the results were utilized to validate the outcome of numerical simulations: (i) Vary the particle fraction of PEG:ChS in a salt-free, neutral bath (H$_b=10^{-7}$ M), where $\alpha$, the particle fraction of ChS, is: 0.1, 0.2, 0.3, 0.4, 0.5 and 1. This experiment is performed to determine the role of fixed charges. Since the ChS monomer volume and the density are constant, increasing the particle fraction of ChS directly corresponds to increasing its weight fraction. (ii) Vary the bath concentration of the monovalent salt within a range of concentrations, 0.1-1 M, and (iii) repeat experiments (i) and (ii) in a salt-free bath and variable pH for neutral hydrogels (i.e., $\alpha = 0$) as well as for ionic gels (i.e., $\alpha > 0$). The pH is varied in the range 3-8 to ensure that the model accurately captures the local variables (i.e., the volume fractions and the osmolarity), within the physiological range [57]. Further, the above mentioned three sets of experiments are repeated for gels with different cross-link fractions. Gels of varying cross-link density are achieved by changing the total monomer concentration in solution prior to polymerization (e.g., 10 wt \%, 20 wt \%, and 30 wt \% gel). Polymer volume fraction is determined from equilibrium swollen volume and dry polymer volume measurements and compared to the numerically determined value. The initial volume of the dry polymer, V$_i$, was determined from the initial mass, m$_i$, using a weighted density of PEG and ChS, as follows
\beq
V_i = 0.1 \Big( \frac{m_i \cdot \% PEG}{\rho_{PEG}} + \frac{m_i \cdot \% ChS}{\rho_{ChS}} \Big) %+ 0.9\frac{m_i}{\rho_{DI-H_2O}}
\eeq
The mass of the gel at equilibrium was measured and used to determine the swollen volume, V$_{swollen}$. It was assumed that all polymer was incorporated in the gel, which was confirmed for chondroitin sulfate, and therefore a change in mass was solely due to the a change in the water content. Hence, for polymers with a 10 wt \% dry weight in solvent,
\beq
V_{swollen} = 0.1 \Big( \frac{m_i \cdot \% PEG}{\rho_{PEG}} + \frac{m_i \cdot \% ChS}{\rho_{ChS}} \Big) + \frac{m_{swollen}-0.1 m_i}{\rho_{solvent}}
\eeq
The initial mass, m$_i$, was the mass of the gel immediately following polymerization and before being placed in a solvent. The swollen mass, m$_{swollen}$, is the mass of the gel at equilibrium. The volume fraction of the polymer was calculated as the volume of the dry polymer (i.e. before swelling) divided by the volume of the swollen gel, $V_i / V_{swollen}$. The kinetic chains associated with methacrylate group (-MA) represents less than one percent of the total gel volume and therefore are not considered in the volume calculations.

%%%%%%%%%%%%%%%%%%%%%%%%%%%%%%%%%%%%%%%%%%%%%%%%%%
\subsubsection{Comparison of model outcomes with experiments: parameter estimation} \label{subsec:parameter}

The values of the parameters used in numerical calculations are listed in Table {\bf 9.1}.  The constants in the model are the monomer volumes, $\nu_1, \nu_2$, the coordination number of the PEG-ChS polymer lattice, $z$, and the nearest neighbor interaction energies, $\epsilon_i$ (\S {\bf 9.5.2}). The undetermined constants are the binding affinities of the polymer gel with Sodium ($K_n$) and hydrogen ($K_n$ ).
\begin{table}
\centering
\begin{tabular}{|c|c|c|c|c|}
\hline
 & PEG ($i=1$) & ChS ($i=2$) & Units \\
\hline
Density ($\rho_i$) & 1.07 & 1.001 & g/mL \\
\hline
Molecular weight (M$_i$) & 4600 & 48700 & g/mol \\
\hline
Repeat unit per chain ($N_i$) & 102 & 86 & -- \\
\hline
Cross-link fraction ($k_i$) & 0.25, 0.5, 0.75 & 0.25, 0.35, 0.45 & -- \\
\hline
Hildebrand solubility ($\delta_i$) & 17.39 & 5.19 & MPa$^{1/2}$ \\
\hline
\multicolumn{2}{|l|}{Hildebrand solubility for water ($\delta_w$)} & 48.07 & MPa$^{1/2}$ \\\hline
\end{tabular}
\caption*{Table 9.1: Parameters common to all the numerical results. The reference temperature for the solubility parameters is fixed at $T_0=298$K.}\label{tab:Table1}
\end{table}

The monomer volumes are found from the density and the molecular weight information ($\nu_i = M_i/(\rho_i * N_A)$). The cross-linked PEG-ChS matrix has a 3-D configuration, which suggests that we choose the coordination number, $z=6$, which mimicks a 3-D cubic lattice. The standard free energies, ${k}_B T_0 \mu^0_p$ and ${k}_B T_0 \mu^0_s$ and the interaction energies, $\epsilon_i$  are found from the Hildebrand solubility data, $\delta_i$ \cite{Barton1990}. The standard free energy is the energy of all the interactions between the molecule and its neighbors in a pure state that have to be disrupted to remove the molecule from the pure state. %The relation between the standard free energies and the solubility parameters (listed in Table \ref{tab:Table1}) are
%
%\begin{align}
%-{\it k}_B T_0 \mu^0_p (\alpha=0) &= \nu_1\delta^2_1 \nonumber \\
%-{\it k}_B T_0 \mu^0_p (\alpha=1) &= \nu_2\delta^2_2 \nonumber \\
%-{\it k}_B T_0 \mu^0_s &= \nu_w\delta^2_w, \label{eq:HSP}
%\end{align}
%
%where $\nu_w=2 \times 10^{-23}$ cm$^3$ is the volume of 1 molecule of water at reference temperature, $T_0 = 298$K. The negative sign in Eqn. (\ref{eq:HSP}) indicates that ${\it k}_B T_0 \mu^0_p, {\it k}_B T_0 \mu^0_s < 0$, since they are the interaction energies. 

The molecular mass and the density of PEG were fixed at 4600 g/mol and 1.07 g/mL, while that of ChS were selected at 48700 g/mol and 1.001 g/mL, respectively. These values give the monomer volumes of the PEG chains as $\nu_1=7.14 \times 10^{-21}$ cm$^3$, and that of ChS chains as $\nu_2=8.08 \times 10^{-20}$ cm$^3$. The density of the solvent, $\rho_{solvent}$, was assumed to be 1.00 g/ml and the volume of 1 molecule of water at reference temperature, $T_0 = 298$K, is $\nu_w=2 \times 10^{-23}$ cm$^3$. The Hildebrand solubility parameters for pure species (values given in Table~\ref{tab:Table1}) and the monomer volumes are used to calculate the interaction energies, $\epsilon_i$ (i=1,...,4). Using the solubility relations [82], these values are fixed at $\epsilon_1=0.0, \epsilon_2=3.74, \epsilon_3=9.58, \epsilon_4=-56.17$. The reference temperature of the experiments was fixed at $T_0=298$K. The undetermined parameters, namely the binding affinity of the polymer gel with various cations ($K_h, K_n$) are computed by minimizing a nonlinear least-square difference function between the experimentally observed values of the equilibrium volume fraction, $\theta_p$, and the corresponding model output, implemented via the MATLAB non-linear least-square minimization function {\bf lsqnonlin}. These values are found as log$_{10}(K_n)=-2.17$, log$_{10}(K_h)=-3.65$.

\begin{figure}[htbp]
\centering
\subfigure[]{\includegraphics[scale=0.5]{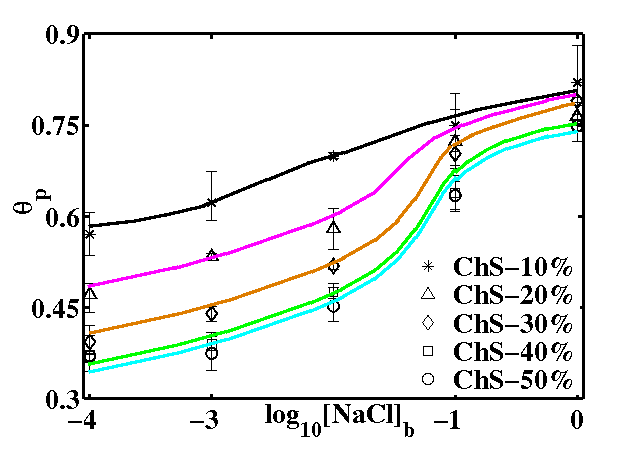}}
\subfigure[]{\includegraphics[scale=0.5]{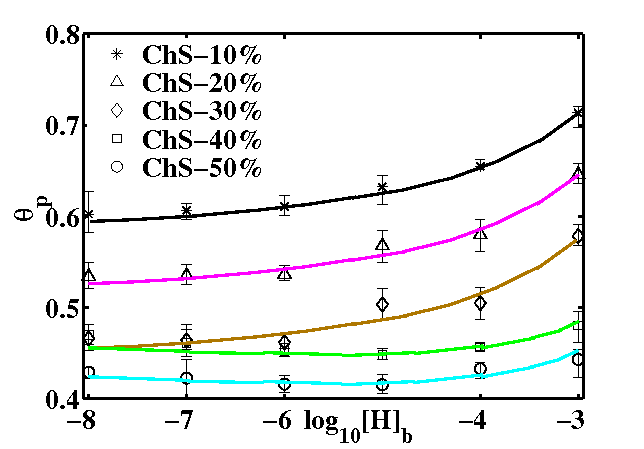}}
\caption*{Fig 9.8: Experimental data of equilibrium volume-fraction of the PEG-ChS gel vs. (a) different NaCl concentration in the bath (in mol/lt) and pH = 7.0, and (b) different pH and zero salt concentration. The gel sample is 10 wt \% in solvent. The cross-link fraction of the PEG/ChS polymer pairs are $k_1=0.25$ and $k_2=0.25$, respectively. Notice the closeness of fit between the model output (i.e., solid lines) and the experimental data points.}\label{fig:Fig6}
\end{figure}

Fig. 9.8 presents the closeness of fit between the model (highlighted by the solid lines) and the sample averaged equilibrium data-points at different salt concentrations in a neutral pH sample (Fig. 9.8a) as well as in a salt-free, variable pH solution (Fig 9.8b). The experimental values were noted at time $t=24$ hrs. Each point represents the average of four samples. The values are noted in identical conditions with the upper and the lower limits in the error bar representing maximum and the minimum variation from the average, respectively.

%%%%%%%%%%%%%%%%%%%%%%%%%%%%%%%%%%%%%%%%%%%%%%%%%%
\subsection{Effect of structural and environmental fluctuations on equilibrium ECM configuration} \label{sec:results}

Articular cartilage is composed of highly negatively charged aggrecan aggegates dispersed in a crosslinked collagen network. This chemical make-up gives rise to an unusually high concentration of mobile-counter ions (e.g., Na+) in the interstitial fluid of the tissue. As a result, the cartilage tissue experiences a complex set of electromechanical cues (signals) arising from fixed negative charges (aggrecan), and mobile-counter ions, in equilibrium. Understanding and recapitulating the electromechanical cues in a tissue engineering approach may be key to bioengineering functionally competent cartilage.

%%%%%%%%%%%%%%%%%%%%%%%%%%%%%%%%%%%%%%%%%%%%%%%
\subsubsection{Effects of changes in ionization, cross-links} \label{subsec:GelCrosslink}

%The first set of numerical experiments were designed to find the relation between the equilibrium configuration and the particle fraction of the charged component of the gel, $\alpha$, of the gel. The experiments are repeated for gels with different cross-link fractions, ($k_1, k_2$). 
Fig. 9a,b,c present the equilibrium volume fraction, corresponding Donnan potential and the net-osmolarity versus the particle fraction of ChS, $\alpha$, respectively. The various curves represent gels with different cross-link fractions, ($k_1, k_2$, Eqn. \ref{eq:rubberElas}). It can be shown that the average charge per monomer (or the average ionization) increases in proportion to the particle fraction, $\alpha$. The gel is dissolved in salt-free, neutral water. Any far field boundary effects are neglected in our numerical simulations, by assuming infinite bath conditions. The negative charges on the gel causes water to dissociate into hydrogen (H$^+$) and hydronium (OH$^-$) ions. The free H$^+$ ions bind with the monomers. 
\begin{figure}[htbp]
\centering
\subfigure[]{\includegraphics[scale=0.5]{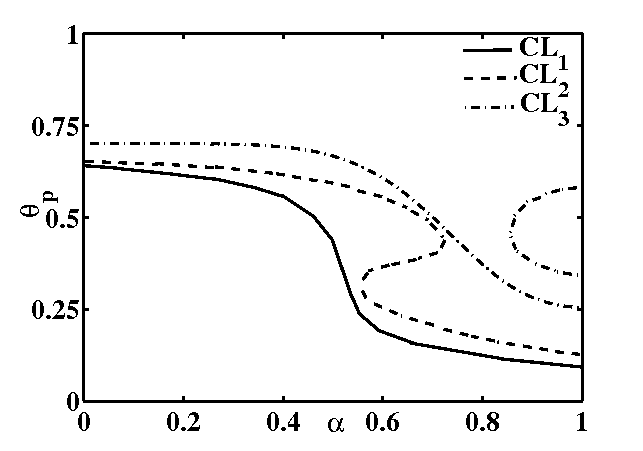}}
\subfigure[]{\includegraphics[scale=0.5]{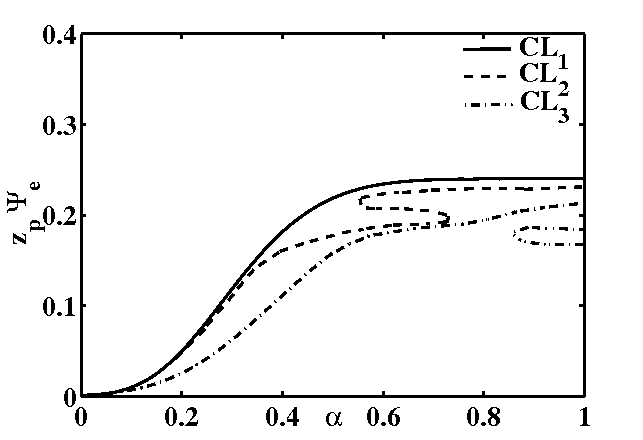}}
\vskip 0.0001cm
\subfigure[]{\includegraphics[scale=0.5]{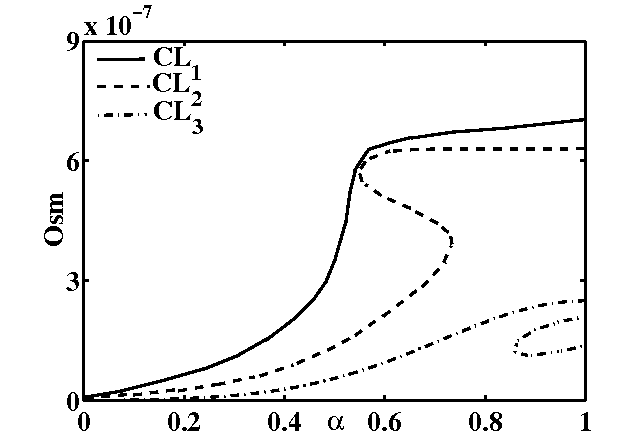}}
\caption*{Fig. 9.9: (a) Equilibrium polymer volume-fraction, (b) Donnan potential, and (c) Net-Osmolarity in neutral salt-free, bath conditions vs. $\alpha$, the particle fraction of chondrotin sulfate component for gel solutions with different 
cross-link fractions.}\label{fig:Fig7}
\end{figure}

At low ionization (or low particle fraction values, $\alpha$), the physically relevant solutions swell as a function of increasing $\alpha$. This is due to the increased swelling effect via the Donnan potential (e.g., look at the trend of the volume fraction and the Donnan potential in the range, $0 \le \alpha \le 0.9$). The physically relevant solutions are the largest and smallest ones, if there are three solutions, because the intermediate solution is unstable. Another trend is that the gel deswells versus increasing $\alpha$, at high cross-link fraction and particle fraction values (the curve denoted by CL$_3$ in the range $\alpha > 0.9$, figure). This observation is corroborated by experimental results that a highly cross-linked network provides a barrier against the Donnan potential and deswells at higher ionization levels [48,67].

%%%%%%%%%%%%%%%%%%%%%%%%%%%%%%%%%%%%%%%%%%%%%%%
\subsubsection{Effects of changes in the bath salt concentration} \label{subsec:SC}
%

%Next we numerically investigate the equilibrium solutions when the gel is immersed in an infinite bath with fixed hydrogen concentration ($H_b = 10^{-7}$ M) and containing a monovalent salt (i.e., a salt which furnishes monovalent cations when dissolved in water, e.g., Na$^+$). The simultaneous binding reactions of the negatively charged gel with two cations (i.e., H$^+$, Na$^+$) are listed in Eqn. (\ref{eqn:chem}). 

Figure 9.10 a,b,c highlight the corresponding equilibrium volume fraction, Donnan potential and the net-osmolarity versus the bath salt concentration, [NaCl]$_b$, for polymer gels immersed in a neutral water solution, respectively. These gels have variable cross-link fractions but a fixed ChS particle fraction, $\alpha=0.1$. Usual features emerge at high concentrations ([NaCl]$_b \ge 10^{-3}$ M) and low concentrations ([NaCl]$_b \le 10^{-8}$ M) of salt. At higher salt concentrations, the gel deswells and the deswelling is more prominent for specimen with greater cross-link fraction. This is readily understood since at higher salt concentrations, more dissolved Na$^+$ ions are furnished which can bind with the negatively charged gel. As a result, the primary ingredient of swelling pressure, i.e., the Donnan potential, is negligible (Fig. 9.10b). High cross-link density within the gel resists the swelling pressure which further enhances the deswelled state of the gel. At lower concentrations of salt, there any hardly any Na$^+$ (or H$^+$) ions readily available to bind with the gel. The Donnan potential is nearly constant and this feature leads to a fixed equilibrium state of the gel ($\theta_p \approx 0.65$).
\begin{figure}[htbp]
\centering
\subfigure[]{\includegraphics[scale=0.5]{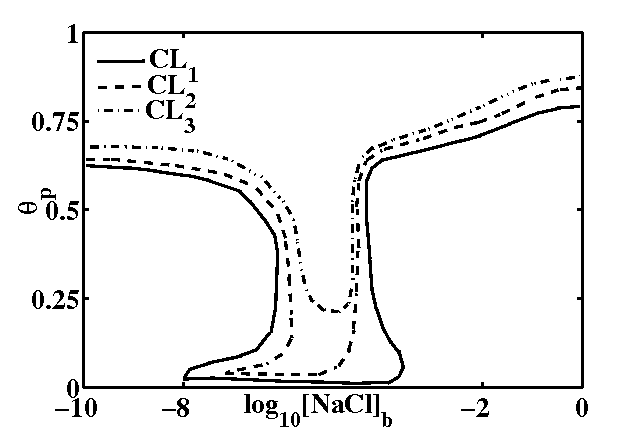}}
\subfigure[]{\includegraphics[scale=0.5]{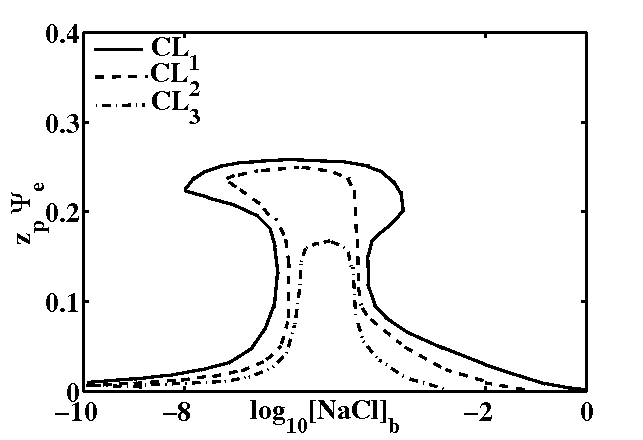}}
\vskip 0.00001cm
\subfigure[]{\includegraphics[scale=0.5]{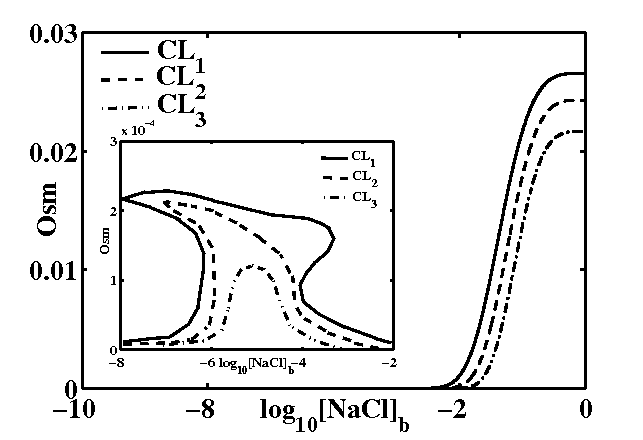}}
\caption*{Fig. 9.10: (a) Equilibrium polymer volume-fraction, (b) Donnan potential, and (c) Net-Osmolarity, vs. NaCl-concentration (in mol/lt.) in the bath. Different curves represent samples a fixed ChS particle fraction, $\alpha=0.1$, and variable cross-link fractions. Note the match in the equilibrium volume-fraction and the experimental data (Fig. 9.8a) in the range, $-4 \le \log_{10}[NaCl]_b \le 0$.}\label{fig:Fig8}
\end{figure}

For intermediate salt concentrations ($10^{-8} < $ [NaCl]$_b < 10^{-3}$ M), a new swelling-deswelling feature emerges, namely, there is a swelling, either gradual or through a phase transition, as the salt concentration in the bath increases. The explanation for these swelling feature is that there is a complicated interplay between ionization (which promotes swelling) and increases in bath concentration of salt which promote deswelling [82].

%%%%%%%%%%%%%%%%%%%%%%%%%%%%%%%%%%%%%%%%%%%%%%%
\subsubsection{Effects of changes in the bath pH} \label{subsec:pH}

Two sets of numerical experiments are performed to predict the equilibrium volume fraction of the charged gel. In the first experiment, cross-link fraction of the polymer matrix is fixed, $k_1=k_2=0.25$, while the ChS particle fraction is varied. Overall, it is found that increasing the bath concentration of hydrogen leads to deswelling. The reason is that a decrease in pH leads to an increase in the dissolved cations which readily bind with the gel, thereby decreasing the gel ionization and the swelling effect via Donnan potential. In this case, swelling is predominantly driven by the ionization which is induced by the Donnan potential. The pressure created by the Donnan potential is about an order of magnitude greater than potential induced osmotic pressure (e.g., compare the order of magnitude of the curves in Fig. 9.11b versus Fig. 9.11c).

Another set of numerical simulations is carried to determine the equilibrium solutions of gels having a fixed ChS particle fraction but a variable cross-link fraction. It is found that gels progressively preferred a deswelled state in the order of increasing cross-link fraction values, consistent with our observations in {\bf 9.5.2}. In summary, for the values of parameters examined, the gel swells predominantly due to effects of increasing Donnan potential and deswells due to the combination of increasing bath concentration of cations and increasing covalent cross-links within the polymer matrix.
\begin{figure}[htbp]
\centering
\subfigure[]{\includegraphics[scale=0.5]{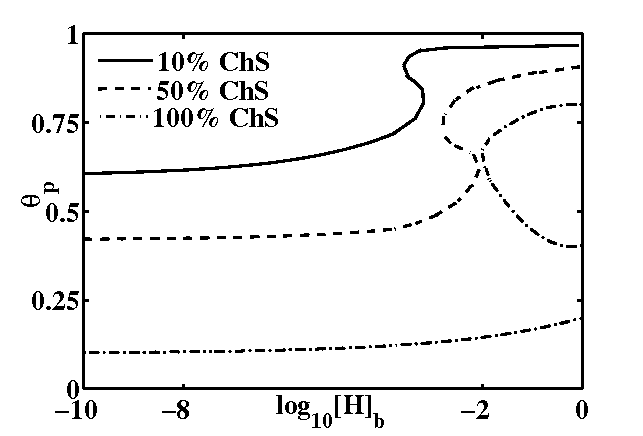}}
\subfigure[]{\includegraphics[scale=0.5]{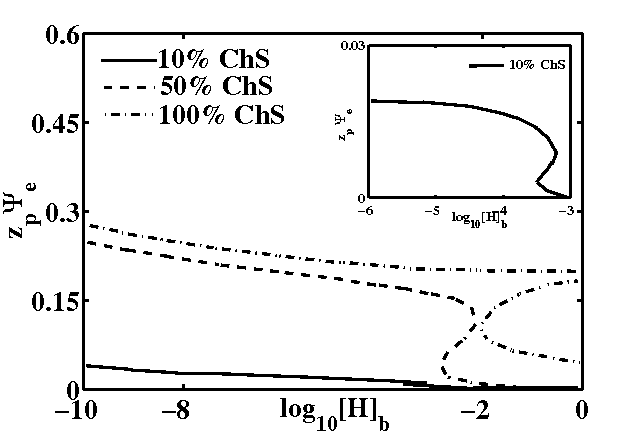}}
\vskip 0.00001cm
\subfigure[]{\includegraphics[scale=0.5]{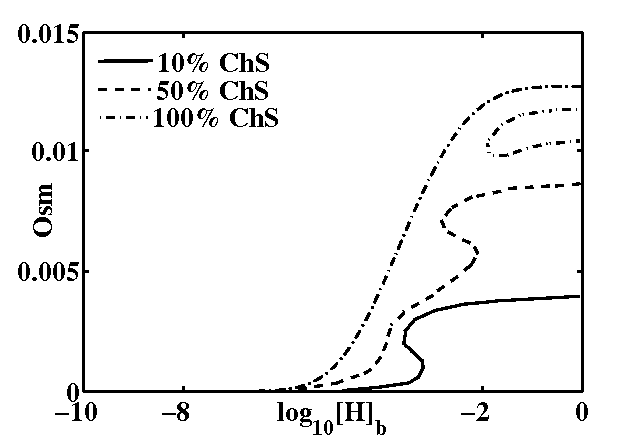}}
\caption*{Fig. 9.11: (a) Equilibrium polymer volume-fraction, (b) Donnan potential, and (c) Net-Osmolarity vs. hydrogen bath concentration $H_b$, for a gel with fixed cross-link fraction $k_1=k_2=0.25$ and variable composition of ChS. Note the match in the equilibrium volume-fraction and the experimental data (Fig. 9.8a) in the range, $-8 \le \log_{10}[H]_b \le -3$.}\label{fig:Fig9}
\end{figure}
%

%%%%%%%%%%%%%%%%%%%%%%%%%%%%%%%%%%%%%%%%%%%%%%%%%%
\subsection{Concluding remarks} \label{sec:conclusion}

In this chapter a description of the makeup of the cartilage tissues is given. We then provided a multi-phase, multi-species model to shed light on the swelling / de-swelling mechanism for ionic gels with covalent cross-links; as well as provided a detailed description of the experimental methods to engineer these biogels. These polymeric biogels mimic the electro-chemical environment of articular cartilage. Finally, we have highlighted the effect of structural and environmental ßuctuations on the equilibrium tissue configuration.

Current research includes determining the counter-ion concentration and local osmolarity in dynamically loaded gels, determining the effect of the environmental fluctuations on the cartilage cell response and the tissue development. A synchronous development of biomechanical models and numerical schemes are following. Readers are requested to look at the research work done by Stephanie Bryant {\it et al.} [13,95] on the experimental side and those by Van Mou {\it et al.} [51,65,66] on the numerical end. Ultimately, in the long term, the scientists and the tissue engineers expect to answer the fundamental question through research in this area, {\em how important are electro-mechano-chemical cues in guiding cellular response and cartilage tissue growth?} Information gleaned from models and experiments will help understand the pathway to cartilage degeneration, repair strategies, maintaining healthy joints via delayed tissue aging and developing superior mechano-chemically resistant biogels as joint replacements.

\end{document}